\documentstyle[epsfig,12pt]{article}
\begin{document}

\newcommand \be  {\begin{equation}}
\newcommand \bea {\begin{eqnarray} \nonumber }
\newcommand \ee  {\end{equation}}
\newcommand \eea {\end{eqnarray}}

\title{{\bf Separation of time and length scales in spin-glasses: temperature 
as a microscope}}

\author{Jean-Philippe Bouchaud$^{1}$, Vincent Dupuis $^1$, Jacques Hammann$^1$,\\
and Eric Vincent$^1$}

\date{\it
$^1$ Service de Physique de l'\'Etat Condens\'e,
 Centre d'\'etudes de Saclay, \\ Orme des Merisiers, 
91191 Gif-sur-Yvette Cedex, France \\ 
}
\maketitle

\begin{abstract}
We summarize the different puzzles raised by aging experiments 
of spin-glasses and their various interpretations. We try to reconcile 
the `real space', droplet
like pictures with the hierarchical pictures that have been proposed 
in the past. The
basic ingredient is a strong separation 
of the time scales that govern the dynamics of the system on different length scales. Changing
the temperature changes the length scale at which the system is observed, thereby allowing 
rejuvenation (that concerns short length scales) and memory (stored in long length scales) to
coexist. We show that previous experiments can be reanalyzed in terms of {\it vanishing
energy barriers} at the spin-glass transition, an important ingredient to obtain a fast separation
of time scales. We propose to distinguish between `fixed landscape
rejuvenation', which is already present in simple two 
(or multi) level systems, 
from the `strong' chaos effect on scales larger than an `overlap length' 
conjectured in the context of the droplet 
model. We argue that most experiments can be accounted for without invoking 
the existence of an overlap length.
New experiments are presented to test some recent predictions of 
the strong chaos scenario, with 
negative results.
\end{abstract}

{\it La complexit\'e de l'ensemble fait que tout ce qui peut leur arriver 
est vraiment, malgr\'e l'exp\'erience acquise, impossible \`a pr\'evoir, 
encore plus \`a imaginer. Il est inutile de tenter de le
d\'ecrire, car on peut concevoir n'importe quelle solution.}

{\sc boris vian}, in {\it L'Automne \`a P\'ekin}.

\section{Facts and puzzles}

Although spin-glasses are totally useless pieces of material, they constitute an exceptionally
convenient laboratory frame for theoretical and experimental investigations \cite{RevSwedes,RevSitges}.
Theoretical concepts and experimental protocols relevant for more `useful' glassy
materials (polymer and molecular glasses, foams and pastes, etc.\cite{SGR}) have been elaborated
and tested on spin-glasses \cite{Revus}. There are at least two reasons for this: (a) the theoretical models
are conceptually simpler (although still highly non trivial) and (b) the use of very
sensitive magnetic detectors allows one to probe in details the a.c and d.c spin dynamics 
of these systems down to very small external fields. The corresponding mechanical measurements in 
other glassy systems are much more difficult to control, although some recent progress have
been made \cite{Derec,Cloitre}, in particular concerning the measurement of age-dependent structure factors
\cite{Cipelliti,Lequeux,Abou}.

The aging dynamics of spin glasses has therefore been studied in glory details recently,
and has revealed an extremely rich phenomenology \cite{RevSwedes,RevSitges,RejMemory,RevJapan}. 
The most striking aspect is the role 
of small temperature changes, that we summarize as follows: 
(a) {\it Superactivated behaviour}: 
time scales grow faster than what would be expected from simple thermal activation when the
temperature is decreased; (b) {\it Rejuvenation and memory}: after a small negative 
temperature jump (within the glass phase) the system behaves as if it had been quenched 
from above the glass transition temperature $T_c$ (rejuvenation). However, 
a perfect memory of the
time spent at the initial temperature is somehow kept, as clearly demonstrated by the now
well known `dip' imprinting experiments (see Fig. 1) \cite{RejMemory}; and (c) 
{\it Weak cooling rate dependence}: the
a.c. susceptibility hardly depends on the thermal history, or actually only on the cooling
rate over the very last Kelvins before reaching the final temperature. The cooling rate 
at higher temperature, in particular when crossing $T_c$, is irrelevant, in strong contrast
with Random Field like systems \cite{DSF,Alberici}. This absence of cooling rate dependence 
is in fact another manifestation of rejuvenation.

\begin{figure}
\hspace*{+1cm}\epsfig{file=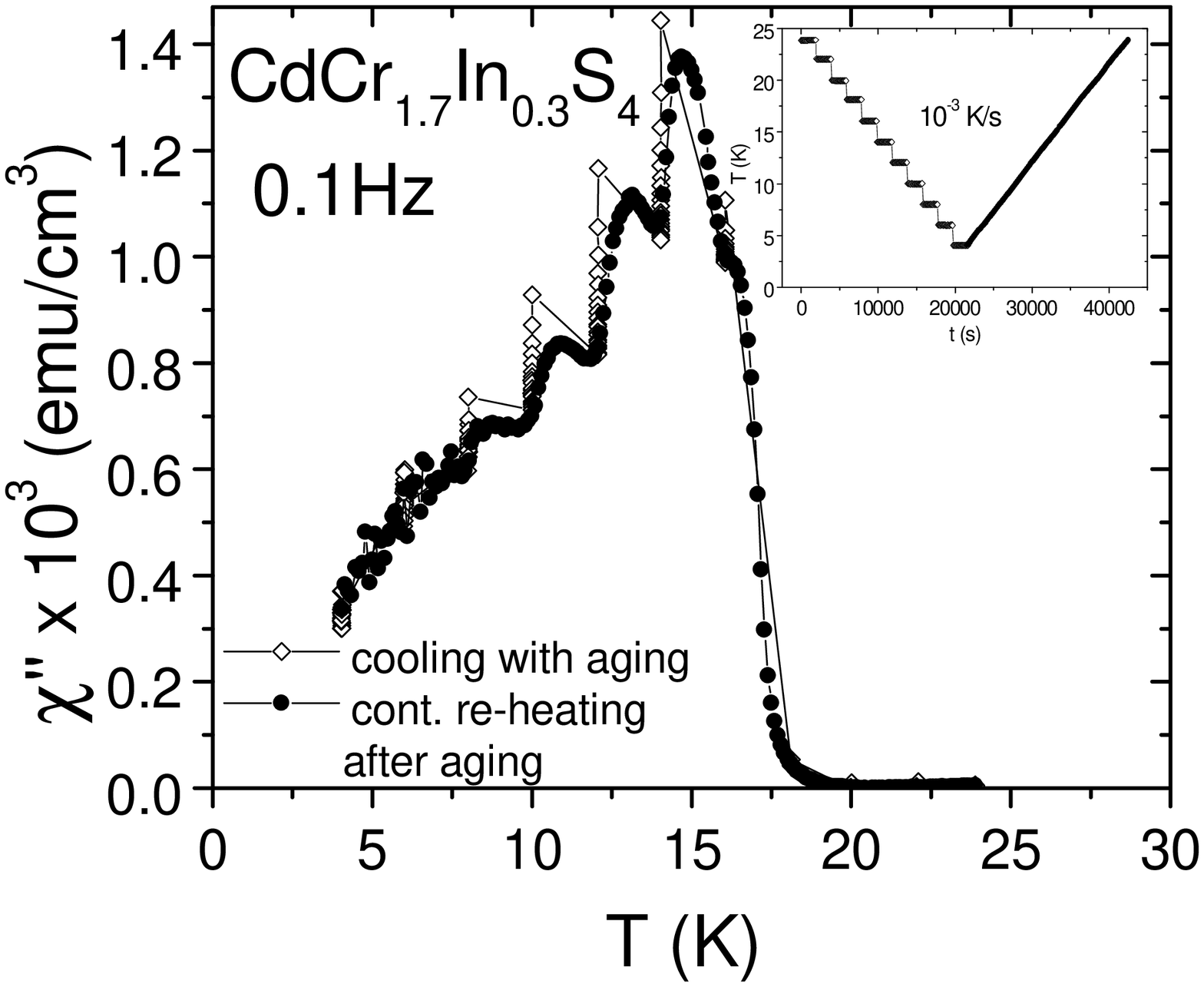,width=10cm}
\vskip 0.3cm \caption{\small Series of `dip' imprinted on the a.c. susceptibility 
by successive stops at different 
temperatures while the system is cooled. Further cooling `rejuvenates' the system
(i.e the susceptibility goes up). However, the dips are one by one remembered by the system
when heated back. For more details, see \protect\cite{RejMemory}.  
\label{fig1} }
\end{figure}

It is now well established, both experimentally \cite{Joh,Takayamaexp}
and numerically \cite{Huse:92,Rieger,Yoshinoell,Marinari} that a certain `coherence' 
length is growing in an aging spin-glass.
This was first
predicted in the context of the `droplet model' \cite{FH}, but is presumably of much more
general validity: larger length scales take a longer time to evolve. This is expected to
be true even if the basic tenets of the droplet model turn out to be incorrect. 
The coherence length is found to grow as a power of time:  $\ell \sim t^{1/z}$, with an apparent exponent 
$1/z$ linear in temperature \cite{Rieger,Yoshinoell,Marinari,Joh}. 
This suggests an activated behaviour over barriers that
grow as the logarithm of the length $\ell$. Indeed, writing $t(\ell) \sim \exp[\Delta(\ell)/k_B T]$ with
$\Delta(\ell)=\Delta_0 \log \ell$ leads to $1/z=k_B T/\Delta_0$. This is
confusing because (a) as mentioned above, experiments suggest superactivated behaviour; 
(b) barriers should grow with $\ell$ faster than excitations energies, which are
thought to grow as $\ell^\theta$ with $\theta=0.2$ in the droplet model. (Note however that recent 
numerical simulations suggest that non compact excitations indeed correspond to $\theta=0$ 
\cite{Martin,Lamarcq}); and 
(c) the exponent $z$ determined numerically or experimentally can be written as $z(T)=z_c T_c/T$,
where $z_c \sim 6$ is the critical exponent that governs the dynamics of the spin-glass at the 
critical point. This coincidence suggests that the system is somehow affected by critical fluctuations
(as also suggested, on the basis of different arguments, in \cite{Bockil}). 

The rejuvenation effect (and absence of cooling rate dependence) brings still more confusion.
Activated barrier crossing is obviously easier at higher temperature: so why waiting longer 
around $T_c$ does not help to equilibrate the system, as happens in e.g. random field like
systems? (see the e.g. the discussion in \cite{Alberici}).
This should be even more crucial if the dynamics is superactivated. 
A way out of this contradiction is
to invoke chaos in temperature \cite{BM,FH,Miya}: 
if new patterns need to equilibrate when the temperature is
changed, it is clear that the time spent at a higher temperature does not help much to
equilibrate the system at the final temperature. However (a) no sign of chaos has been 
found in most recent static numerical studies of the 3d Edwards-Anderson model \cite{Billoire} (at variance with earlier studies \cite{Nifle}), nor in
the theoretical analysis of the {\sc sk} model \cite{Rizzo};
(b) no rejuvenation in the dynamics upon temperature changes has been found either in numerical 
simulations \cite{Ritort}; and
(c) the coherence length that has grown at one given temperature seems to carry on growing 
(although at a different rate) at another temperature, in contradiction with the chaos idea. 
This continuity has been established both numerically 
\cite{Yoshinoell} and experimentally \cite{Takayamaexp}. 
`Chaos' should furthermore be compatible with memory: growing new patterns at a given
temperature should not erase the patterns grown at higher temperatures if one wants to account 
for the `multi-dip' experiment shown in Fig. 1. A possible scenario for this was suggested in 
\cite{Yoshino/Lemaitre:2000}.

In this paper, we wish to develop a consistent qualitative picture for the dynamics in
spin-glasses that allows one to resolve the above apparent contradictions. This picture,
as the original droplet model \cite{KH,FH}, heavily relies on three basic ideas: 
(i) times grow as the exponential of the energy barriers; (ii) the energy barriers grow 
as a power of the length scales involved in the dynamics; and (iii) the energy barriers 
vanish at the critical temperature $T_c$. These ingredients are enough to understand that
changing the temperature for a given observation time corresponds to changing the length
scale at which the system is probed. Therefore, the objects contributing to the aging
dynamics are different at different temperatures. This gives a precise content to the
`hierarchical' picture advocated in many papers \cite{Vincent,Lefloch,Weissmann,Bouchaud/Dean:1995}.
Although the present picture is 
similar in spirit to the droplet model, we will assume (but this is not crucial) that 
the low energy metastable states exist on arbitrary length scales (i.e. $\theta=0$). 
We will also introduce the idea of a fixed landscape rejuvenation, which concerns short length 
scales and is distinct from the large scale chaos that appears in the context of the droplet model.
We reanalyze previous experimental data within this framework, which strongly suggest
that energy barriers vanish at $T_c$, and allow us to extract 
estimates of the exponent $\psi$ that relates energy barriers and length scales.

\section{Basic ingredients}

Let us consider a large scale low lying excitation in a spin glass. This excitation is made
of a large connected cluster of spins that is flipped with respect to the ground state. The surface
of this cluster can be thought of as a `domain wall' which happens to occupy a very favorable 
position since the overall energy of this surface is very small. Therefore, this wall is `pinned' 
by the disorder and tends to adopt some special conformation. If $\theta=0$, this wall has no 
overall tendency to disappear with time, and will be present in the system even after very long times.
If $\theta > 0$, conversely, these large scale walls tend to fraction in smaller and smaller bubbles
before disappearing in the equilibrium state. However, in some {\it exceptional} circumstances, the
energy of these droplets is smaller than $k_B T$, and these walls survive in equilibrium. In the droplet
picture, this occurs with probability $k_B T/\ell^\theta$ \cite{FH}. 

Now, there are many conformations of the domain wall which have approximately the same energy. One
can flip clusters of spins that touch the domain wall at a small cost, corresponding to a local
modification (`blister') of the conformation of the wall. These excitations can occur on all length scales:
one can create blisters within blisters, etc. The situation here is not specific to spin-glasses but
is also true for a domain wall in a disordered ferromagnet that can adopt many different metastable 
configuration. An important difference is that in a disordered ferromagnet, these domain walls have 
a positive energy and tend to disappear with time: this is the coarsening phenomenon. The evolution of the
system is in this case a slow but irreversible march towards order \cite{Vincent-Walls,Levelut:2001}. 
The evolution in phase space is biased by the fact that these walls cost energy. 

It is useful to decompose the conformations of these pinned `domain walls' on different length scales.
Identical conformations on length scale $\ell_n$ may differ by the presence or absence of blisters of 
smaller sizes $\ell_{n-1},\ell_{n-2},...$, where $\ell_n = b^n \ell_0$, and $b$ an arbitrary factor,
say $b=2$. Here again, we follow ideas developed in the context of the droplet model for spin-glasses 
\cite{FH}
or pinned domain walls \cite{Nattermann,FH2}. 
In the droplet model, the time needed to evolve the conformation on scale $\ell_n$ is
taken to be:
\be
t_n = t(\ell_n) \sim \tau_0 \exp\left(\frac{\Upsilon \ell_n^\psi}{k_B T}\right)\label{FHtime}
\ee 
where $\tau_0$ is a microscopic time, $\Upsilon$ a typical energy setting the scale of 
energy barriers between conformations, and $\psi$ the so-called barrier exponent. This form
has an immediate consequence: the time needed to evolve the
system on scale $\ell_n$ is extremely long compared to the time needed to evolve the system
on scales $\ell_{n-1},\ell_{n-2},...$. This means, 
as emphasized in \cite{StAndrews}, that on a time scale $t_n$, 
all excitations on 
scales $\ell_{n'}$ with $n'>n$ are essentially {\it frozen}, whereas all excitations on scales
$\ell_{n'}$ with $n'<n$ are essentially {\it equilibrated}. Note also that short
length scales are `slaved' to large length scales: when a large length scale flips over, all the
smaller length scales have to re-equilibrate in a new environment. In this sense, the dynamics is
hierarchical (see a related discussion in \cite{Weissmann}. 

We will choose to write Eq.(\ref{FHtime}) in a generalized form, more appropriate to describe the
vicinity of the spin-glass transition\footnote{In the following, all length scales are expressed in 
lattice size units.}:
\be
t_n = t(\ell_n,T) \sim \tau_0 \ell_n^{z_c} \exp\left(\frac{\Upsilon(T)\ell_n^\psi}{k_B T}\right)\label{tcrit},
\ee 
with $\Upsilon(T)=\Upsilon_0[T_c-T/T_c]^{\psi \nu}$, $\nu$ being the critical exponent governing the
divergence of the equilibrium correlation length $\xi(T)$ at $T_c$: $\xi(T)=|T_c-T|^{-\nu}$. Therefore, 
the term in the exponential can be rewritten as: $(\Upsilon_0/k_B T) [\ell_n/\xi(T)]^\psi$. 
Since $\Upsilon_0$
is expected to be of the order of $k_B T_c$, one sees that as long as $\ell_n$ is {\it smaller} 
than $\xi(T)$, barriers are small compared to $k_B T$
and the exponential term in Eq. (\ref{tcrit}) can be set to $1$. 
This leads to the usual critical dynamics (non activated) relation:
\be
t_n \sim \tau_0  \ell_n^{z_c}.
\ee
Conversely, for length scales larger than $\xi$, the exponential becomes the dominant factor, 
and one recovers 
(to logarithmic accuracy) the Fisher-Huse relation (\ref{FHtime}) with the microscopic time 
$\tau_0$ replaced by the typical critical
time scale $\tau_0 \xi^{z_c}$. As we shall find below, the experiments are typically in 
a crossover regime 
where barriers are larger than, but comparable to $k_B T$. Therefore, we keep the full form of Eq. 
(\ref{tcrit}), which correctly interpolates between the two regimes, to describe the intermediate
regime. 

\begin{figure}
\hspace*{+1cm}\epsfig{file=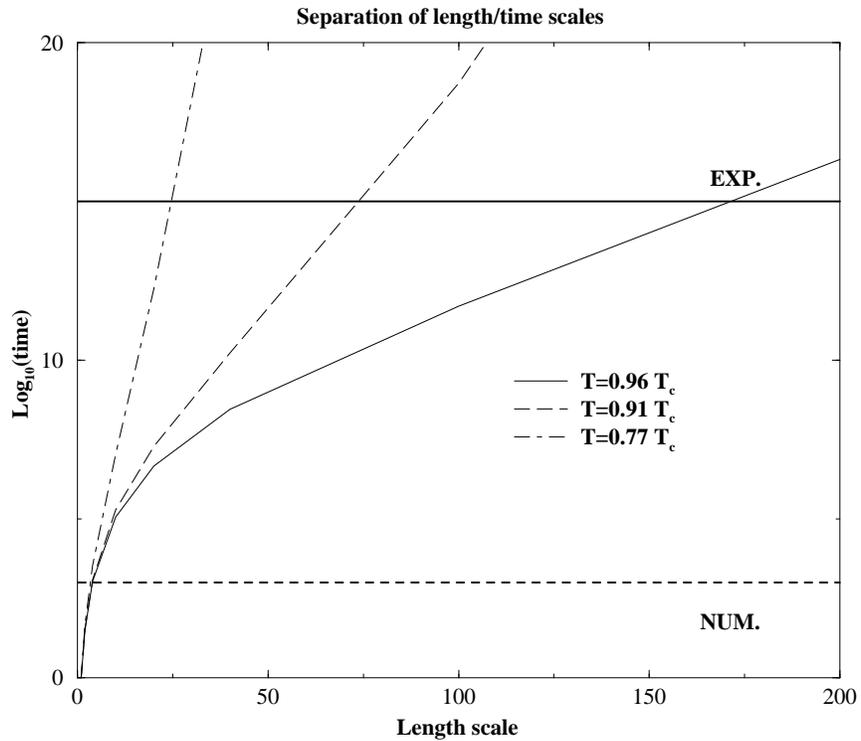,width=10cm,angle=270}
\vskip 0.3cm \caption{\small Logarithm (base 10) of equilibration time versus length scales, 
as given by Eq. \protect\ref{tcrit}, for three different temperatures, corresponding to the experimental data
of Table 1. The thick horizontal line corresponds to 1000 sec. ($10^{15}$ in microscopic
units), and the dotted horizontal line to typical numerical time scales. One should notice that
for experimental time scales, length scales are rather modest, but do separate when the temperature
is changed, at variance with numerical simulations.  
\label{fig2} }
\end{figure}

Several interesting consequences of Eq. (\ref{tcrit}) are worth discussing. 
\begin{itemize}
\item Fig. 2 shows $\log_{10} t_n$ versus $\ell_n$ for a choice of parameters suggested by the 
experiments on AgMn: $\psi=1.5$, $\Upsilon_0/k_B T_c=2.$, $z_c=5$ and $\nu=1.3$, and for different values of
$T/T_c$. 
The thick horizontal line corresponds to $t/\tau_0 = 10^{15}$, corresponding to an experimental 
time scale of $1000$ seconds. 
One sees that the associated length scales are very modest, in the range 10 to
100, and change appreciably when the temperature is changed only slightly. The horizontal dotted line
corresponds to numerically accessible time scales. 
We see that in that case, length scales are extremely
small and do not separate at all with temperature.  
\item Let us fix a certain length scale $\ell$ and change the temperature from $T_1$ to $T_2=T_1-\Delta T$.
The associated time scale $t_{n1}$ at temperature $T_1$ is changed into $t_{n2}$ given by:
\be
\frac{t_{n2}}{\tau_\ell} =\left(\frac{t_{n1}}{\tau_\ell}\right)^\beta \qquad \beta=\frac{T_1}{T_2}
\left(\frac{T_c-T_2}{T_c-T_1}\right)^{\psi \nu}
\ee
with $\tau_\ell = \tau_0 \ell^{z_c}$. This leads to superactivation effects.
For example, with $T_1=0.9 T_c$ and $T_2=0.8 T_c$, $\psi \nu=2$,
one finds $\beta \sim 4$ ! As soon as  $t_{n1} \gg \tau_\ell$, the value of $t_{n2}$ is 
astronomically large.
This means that the separation of time scales when the temperature is lowered is extremely fast, and
to a good approximation, those length scales that are aging at temperature $T_1$ become 
completely frozen at
a slightly lower temperature. In this sense, temperature acts as a microscope. 
If $\Upsilon$ was temperature independent, one would find 
$\beta=T_1/T_2 = 1.125$ for the above choice of parameters. The separation of time scales would then 
only be mild.
\item If one fits Eq. (\ref{tcrit}) with a temperature dependent power law over a restricted 
range of times, one finds an effective exponent:
\be
z_{eff} = \frac{d \log t_n}{d\log \ell_n} = z_c + \psi \frac{\Upsilon(T)\ell_n^\psi}{k_B T},
\label{zeff}
\ee
that grows when the temperature is lowered. We will come back to this point later.
\end{itemize}

\section{Old experiments revisited}

\subsection{Small temperature jumps}

We now revisit two sets of experiments and show that the results are indeed compatible with 
Eq. (\ref{tcrit}). The first set of experiments concerns {\sc trm} relaxation. One first measures an
array of curves corresponding to the standard protocol: cool the system from above $T_c$ to $T_1$, 
leave a (small) magnetic field on for a certain waiting time $t_{w1}$, and observe the relaxation at time
$t_{w1}+t$. One finds curves that scale approximately as $t/t_{w1}$ once the stationary (fast) 
initial part
has been subtracted. Then, a second protocol is followed: cool now to $T_2 = T_1 -\Delta T$, and 
wait for a time $t_{w2}$. Then heat up the system to $T_1$ and simultaneously cut the field. What 
effective waiting time $t_{w1}^{eff}$ should one choose to match this second type of {\sc trm} with a 
standard, isothermal one ? The experimental results show that as long as $\Delta T$ is small enough, 
it is always possible to find such a $t_{w1}^{eff}$ so that the curves match perfectly. This means
that the objects involved in the dynamics are exactly the same for the two temperatures. For larger
$\Delta T$'s, the curves are distorted and such a perfect matching is impossible. Typically,
experiments were performed with very small $\Delta T=0.02$ K ($\Delta T/T_g \sim 0.2 \%$). Obviously, 
since the time spent at $T_2$ corresponds to a smaller time spend at $T_1$, one finds 
$t_{w1}^{eff}< t_{w2}$, see Table. It was furthermore shown in \cite{Hammann} that the correspondence 
between $t_{w1}^{eff}$ and $t_{w2}$ could not be understood in terms of simple thermal activation.
More precisely, one finds that $\log t_{w1}^{eff}/\tau_0 < (T_2/T_1) \log t_{w2}/\tau_0$, corresponding to 
superactivation (unless $\tau_0$ is chosen to be unphysically small). This was interpreted in 
\cite{Hammann} as indicating a divergence of the corresponding
barrier at smaller temperatures. Note however that a barrier involving a finite number of
spins cannot diverge at any temperature. Another interpretation is that barriers actually 
{\it vanish}
at $T_c$, as in the droplet model. This is reasonable since the `domain walls' (whatever
their precise nature) become more and more loosely defined and can
no longer be pinned at $T_c$ (this is also true in a disordered ferromagnet -- but not in a random field
system: see \cite{DSF}). 

\begin{table}
\begin{center}
\begin{tabular}{||c|c|c|c||} \hline\hline
$T_1$ (K)\ \hspace{0.1cm} \  
& \hspace{0.1cm} $\Delta T$ (mK) \hspace{0.1cm}
& \hspace{0.1cm} $t_{w2}$ (sec)\hspace{0.1cm} 
& \hspace{0.1cm} $t_{w1}^{eff}$ (sec) \hspace{0.1cm} \\ \hline

10  &     20   &   300    &    170 \\ \hline
10  &     20   &  1000    &    400\\ \hline
10  &     20   &  3000    &   1400\\ \hline 
10  &     20   & 10000    &   3750\\ \hline
9.5  &    20   &   300    &    165\\ \hline
9.5  &    20   &  1000    &    485\\ \hline
9.5  &    20   &  3000    &   1350\\ \hline 
9.5  &    20   & 10000    &   4550\\ \hline
9.5  &    20   &  30000   &   13000\\ \hline
9    &   20    &  1000    &    650\\ \hline
9    &   20    & 3000     &  1900\\ \hline 
9    &   20    & 10000    &   5500\\ \hline
8    &   20    &  300     &   210\\ \hline
8    &   20    &  1000    &    700\\ \hline
8    &   40    &  300     &   180\\ \hline
8    &   40    &  1000    &    480\\ \hline
8    &   40    &  3000    &   1400\\ \hline 
8    &   40    &  10000   &    4400\\ \hline 

\end{tabular}
\end{center}
\caption[]{\small Effective waiting time $t_{w1}^{eff}$ at $T_1$ versus real 
waiting time $t_{w2}$ at $T_2=T_1-\Delta T$ for different initial temperatures.
The sample is Ag Mn, with a spin-glass temperature
of $10.4$ K. From \protect\cite{Hammann}.}
\end{table}

We have therefore 
reanalyzed the very clean experimental data on Ag Mn of ref. \cite{Hammann} 
by postulating Eq. (\ref{tcrit}). We fix $z_c=5$ and 
$\nu=1.3$ 
to reasonable values and determine $\psi$ and $\Upsilon_0/k_B T_c$ such 
that the length scale $\ell_1$ 
corresponding to the time $t_{w1}^{eff}$ and the length scale $\ell_2$ corresponding to the 
associated time $t_{w2}$ are as close as possible: since the shape of the {\sc trm}'s are 
made to coincide,
the corresponding length scales should also coincide. Therefore, we choose 
$\psi$ and $\Upsilon_0/k_B T_c$ such that the mean squared relative difference:
\be
{\cal E}^2=\sum_{i=1}^{N} \left(\frac{\ell_1^i-\ell_2^i}{\ell_2^i}\right)^2,
\ee
summed over all experiments reported in Table 1, is as small as possible. 
The best values are found to be in a `crescent' in the $\psi,\Upsilon$ plane
with for example $\psi=1.5$ and $\Upsilon_0/k_B T_c=2.$, corresponding to 
a rather small root mean square relative error of ${\cal E}=0.48 \%$, or 
$\psi=2$ and $\Upsilon_0/k_B T_c=0.3$, 
corresponding to an error of ${\cal E}=0.40 \%$ (the experimental error on $t_{w1}^{eff}$
corresponds to a relative error on $\ell$ on the order of $0.5 \%$). The typical length scales
obtained are shown in Fig. 2. 
The value $\psi=2$ corresponds with the upper bound proposed by Fisher and Huse for an  Ising
spin glass. A similar 
range of values ($\psi=1.3$ for $\Upsilon_0/k_B T_c=1$) is found for the insulating 
CrIn compound \cite{Ising}, using the same procedure but on less precise data (larger $\Delta T$). 
The point here is not to claim a very good precision on the value of $\psi$, but rather
to show that the results are compatible with the idea that the barriers continuously
vanish at $T_c$. 
Note that more recent experiments on the role of small temperature 
jump on {\it Ising} spin-glasses confirm the present analysis, although the
separation of time scales is much `milder':
the dynamics for $T > 0.6 T_c$ involve both activated events over
temperature dependent barriers and critical dynamics that increases the effective value 
of the `trial time' $\tau_\ell = \tau_0 \ell^{z_c}$. \footnote{The fact that a rather large
value of this trial time was needed to account for the experiments was also noticed in
\cite{Fieldus}.} However, the value of $\psi$ for this Ising-like sample  
is significantly smaller ($\psi \sim 0.3 - 0.5$) than is the less anisotropic sample reported above,
a somewhat counter-intuitive result.

\subsection{A time dependent length probed by magnetic field}

We now turn to another set of more recent experiments \cite{Joh}, which exploits the fact 
that a small magnetic field acts as to reduce the energy barriers, suggested in \cite{Fieldus}. 
If the number of spins involved in a re-conformation is $N$, one can expect 
that a field $H$ will perturb the barriers by an amount proportional to $N \chi H^2$,
where $\chi$ is the magnetic susceptibility. The typical relaxation time of the {\sc trm} 
aged for a time $t_w$ is 
therefore multiplied by $\exp(-\alpha N(t_w,T) \chi H^2/k_B T)$, where $\alpha$ is a
numerical factor. By measuring this reduction 
factor for different waiting times $t_w$, one can estimate the typical number of spins
involved in the dynamics as a function of the waiting time. Finally, by writing 
$N(t_w,T) \propto \ell^3$, one has access to a time dependent coherence length $\ell(t_w,T)$. As
mentioned in the introduction, this dependence can be fitted, for three different types
of spin glasses, by a power law with a
temperature dependent exponent $z(T)$. Note that the estimated number of spins, in the range 
$10^5 - 10^6$, is compatible with the length scales reported in Fig 2.

\begin{figure}
\hspace*{+1cm}\epsfig{file=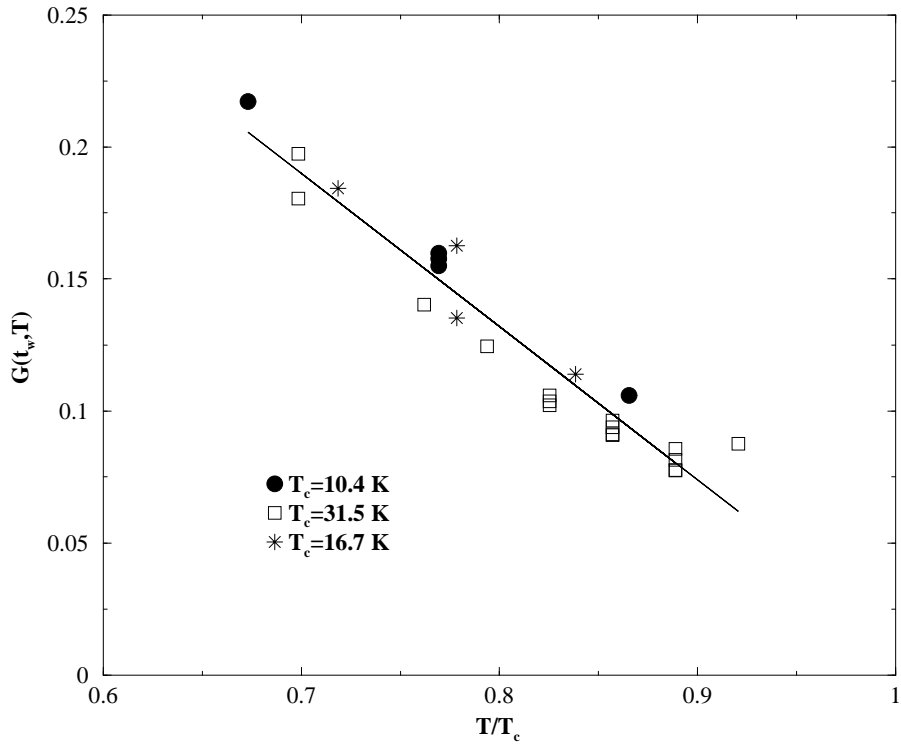,width=10cm,angle=270}
\vskip 0.3cm \caption{\small Plot of the experimentally determined quantity $G(t_w,T)$ for different
waiting times, temperatures, and three different spin-glasses with different $T_c$, plotted as a function of $T/T_c$.
Also shown is a linear regression through all the data points, extrapolating to zero at $T/T_c=1.025$. 
\label{fig3} }
\end{figure}

Here, we want to reanalyze the data of \cite{Joh} in the light of Eq. (\ref{tcrit}), 
and show that
again, these experimental results suggest that barriers vanish at $T_c$. In order to do so, 
we have plotted the quantity:
\be
G(t_w,T)=\left(\frac{\log t_w/\tau_0 - \frac{z_c}{3} \log N(t_w,T)}{\frac{T_c}{T} N(t_w,T)^{\psi/3}}\right)
^{1/\psi\nu}
\ee
as a function of $T/T_c$, for different spin-glasses. If Eq. (\ref{tcrit}) is correct, one
should observe $G(t_w,T)=G_0(1-T/T_c)$, where $G_0$ is a numerical factor. The vanishing of
$G(t_w,T)$ is direct manifestation of the vanishing of the barriers. The results are shown
in Fig 3. We have kept the values $z=5$, $\psi=1.5$ suggested above. A linear fit through the 
points is very reasonable; the most interesting point is that this linear fit is found to
be $G_{fit}(t_w,T)=0.58\ (1.025-T/T_c)$, very close to what is expected from Eq. (\ref{tcrit}).
Note that the extrapolated value of $T/T_c$ is even closer to one if $z_c$ is chosen to be equal to $6$.
Therefore, the power-law dependence of $\ell$ reported in \cite{Joh} (and also in the
numerical work of \cite{Rieger,Yoshinoell,Marinari}) might actually be an effective power law, as suggested by
Eq. (\ref{zeff}) above, which naturally matches the critical dynamics exponent when $T \to T_c$. 

In summary, we have shown in this section that two completely independent sets of experiments
can be interpreted consistently within the framework of section 2: time scales and length scales
are related by Eq. (\ref{tcrit}). The most important aspect is the fact that energy barriers 
vanish at $T_c$. This means that (a) dynamics is superactivated and time scales 
separate extremely fast in spin-glasses, and (b) the critical point does significantly 
affect the dynamics by slowing down the `microscopic' frequency.

\section{Fixed landscape rejuvenation and Memory}

\subsection{Qualitative ideas}

Let us now come back in more details to the temperature cycling experiments of \cite{RejMemory,Levelut:2001}
and show how
these can be qualitatively interpreted within the general picture of section 2. The crucial
observation is that for a given time scale $t_{w1}$ and temperature $T_1$, the aging dynamics is
dominated by a characteristic length $\ell_{1}$ such that $t(\ell_{1},T_1)=t_{w1}$. 
Larger length scales 
are essentially frozen and do not evolve on the time scale of the experiment, while shorter 
length scales are fully equilibrated (for a given larger scale conformation) and only 
contribute to the stationary part of the response function. Note that this picture is obviously a
caricature because the energy barriers in disordered systems are expected to fluctuate in space. Therefore,
some larger length scales might locally see an exceptionally low barrier, and vice versa.

Now, let the temperature change from $T_1$ to $T_2=T_1 -\Delta T$. Because of the separation of time
scales, a relatively small $\Delta T$ is sufficient to freeze completely the dynamics on scale 
$\ell_{1}$ (which will therefore retain the memory of the stay at $T_1$) and to slow down the 
initially fast dynamics on shorter length scales, selecting a particular one 
$\ell_{2}$ to be in the experimental time window. Since at $T_1$ this length scale is equilibrated, the 
different conformations appear in the course of time with their Boltzmann weights. As the temperature is
changed, these Boltzmann weights are modified and the system has to evolve towards a new state. This 
is true even if the (free)-energy landscape does not significantly evolve between the two 
temperatures: this is what we call {\it fixed landscape rejuvenation}. 
Take for example a simple two-level (for example two conformations of a domain wall
differing by a `blister' of scale $\ell$), with an energy difference $E(\ell)$. The population 
difference between the two levels change significantly (say by at least 10\%) if the change of 
temperature is such that:
\be
\Delta T \frac{\partial \tanh(\frac{E(\ell)}{2k_B T})}{\partial T} = \frac{\Delta T E(\ell)}{2k_B T^2} 
\frac{1}{\cosh^2(\frac{E(\ell)}{2k_B T})} \geq 0.1
\ee
The minimal value of $\Delta T$ for such a rearrangement to occur is therefore 
$\Delta T^*/T \sim 0.15$ which is corresponds to $E(\ell) \sim 2.4 k_B T$. (For multi-level
systems, $\Delta T^*$ can be much smaller than this, see below and \cite{Marta}). It is furthermore
easy to show that when a two-level system is driven out-of-equilibrium by a rapid change 
of temperature, an excess low-frequency dissipation follows. The out of phase 
a.c. susceptibility is indeed given by:
\be
\chi''(\omega,t_w) \sim \frac{\exp[-t_w/t(\ell,T_2)]}{\omega t(\ell,T_2)} \qquad \omega t(\ell,T_2) 
\gg 1
\label{2levelchi}
\ee
where $t(\ell)$ is the relaxation time of the two-level system. Therefore, cooling a disordered system can 
induce a strong increase in the out of phase susceptibility (rejuvenation) if there are metastable
states such that $t(\ell_{2},T_2) \sim \omega^{-1}$ and $E(\ell_2) \sim k_B T$. 

Lower frequencies therefore probe larger length scales $\ell_2$. In the droplet model, $E(\ell)$ is typically of order
$\Upsilon \ell^\theta$, and the probability to observe an `active' droplet of energy of order $k_B T$
is $(k_B T/\Upsilon) \ell^{-\theta}$. If $\theta$ is positive, 
we expect that rejuvenation should asymptotically
disappear as $\omega \to 0$ since the probability of observing a large droplets goes to zero. 
However, since $\ell_2$ depends logarithmically on frequency and 
$\theta$ is small ($\leq 0.2$), 
this will in practice never happen and things will look very much as if $\theta=0$.
In other words, the fact that $\ell$ cannot much exceed $100$ in experiments and that $\theta$ is so small,
means that $\ell^{-\theta} \geq 0.4$! The influence of large scale, 
low energy excitations (of order $k_B T$), is therefore dominant in real spin-glasses. This feature 
is actually the basic outcome of mean-field models.

As emphasized above, the strong separation of time scales enables one to observe simultaneously 
rejuvenation and memory if $\Delta T$ is large enough: the length scales that one observes at $T_1$ 
are totally frozen at temperature $T_2$ and therefore resume aging, unaffected by the long stay at 
smaller temperatures. Of course, the states on scale $\ell_2$ are now out of equilibrium at $T_1$; however,
the time needed for them to equilibrate is very short, since it is given by:
\be
t(\ell_2,T_1) = \tau_{\ell_2} \left(\frac{t(\ell_2,T_2)}{\tau_{\ell_2}}\right)^{1/\beta}
= \tau_0^{1-1/\beta} \ell_2^{z_c(1-1/\beta)} \omega^{-1/\beta}
\ee
Taking for example $\beta=2$, $\ell_2=10$, we find that $\omega t(\ell_2,T_1)= 10^{3} (\omega \tau_0)^{1/2}
\sim 10^{-3}$ for $\omega=1$ Hz. Therefore, these length scales have equilibrated far before the
first oscillation of the a.c. field has taken place.

It is interesting to remark that in the scenario, memory cannot be observed after a positive
temperature cycle $T_1 \to T_1 + \Delta T$, if $\Delta T$ is large enough to induce rejuvenation. 
This is because upon heating,
length scales larger than $\ell_1$ will now un-freeze and evolve. Since the shorter length scales are 
slaved to the larger
length scales, the length scale $\ell_1$ will itself have to re-equilibrate completely at
higher temperatures -- this erases the memory of the stay at $T_1$. As will be discussed below, this is not
necessarily true in the `chaos' scenario. Note that the `slaving' of small length scales to
large length scales also explains qualitatively the `second noise spectrum' 
experiments of Weissmann \cite{Weissmann}, which reflect the fact 
that when a large length scale `jumps', all the
smaller length scales have to re-equilibrate in a new environment, uncorrelated with the first
\cite{Weissmann,Bouchaud/Dean:1995}. 

Before describing a more realistic model, let us summarize the above discussion by a schematic 
figure in the plane $\ell,\Delta T$, showing how the different length scales evolve during the
temperature cycle. This figure allows to account qualitatively for the temperature cycling experiments,
in particular the fact that memory is not perfect if $t_{w2}$ is large enough or $\Delta T$ small
enough (see also \cite{Anders}). If $\Delta T<\Delta T^*$, no rejuvenation is expected and aging continues, although 
at a slower rate. Rejuvenation and memory are made possible by the strong length and time scale separation
as a function of temperature for a fixed time scale. As shown in Fig. \ref{fig2} this is not
true in numerical simulations, where one cannot probe the multi-scale dynamics of the system.

Note finally that the above discussion is not restricted to spin-glasses, and can be applied
to describe the aging dynamics of pinned domain walls in ferromagnets \cite{Vincent-Walls},
ferroelectrics \cite{Levelut:2001}, and glassy polymers \cite{pmma} 
where indeed similar effects have been observed experimentally.
There is a significant difference, though, which is due to domain coarsening. The dynamics of the
system progressively gets rid of the domain walls; this corresponds to a {\it cumulative} aging effect,
which is cooling rate dependent. As time evolves, the fraction of spins that belong to domain walls and
contribute to aging dynamics systematically decreases \cite{Berthier}, as observed in these systems 
\cite{Vincent-Walls,Levelut:2001}.  

\begin{figure}
\hspace*{+1cm}\epsfig{file=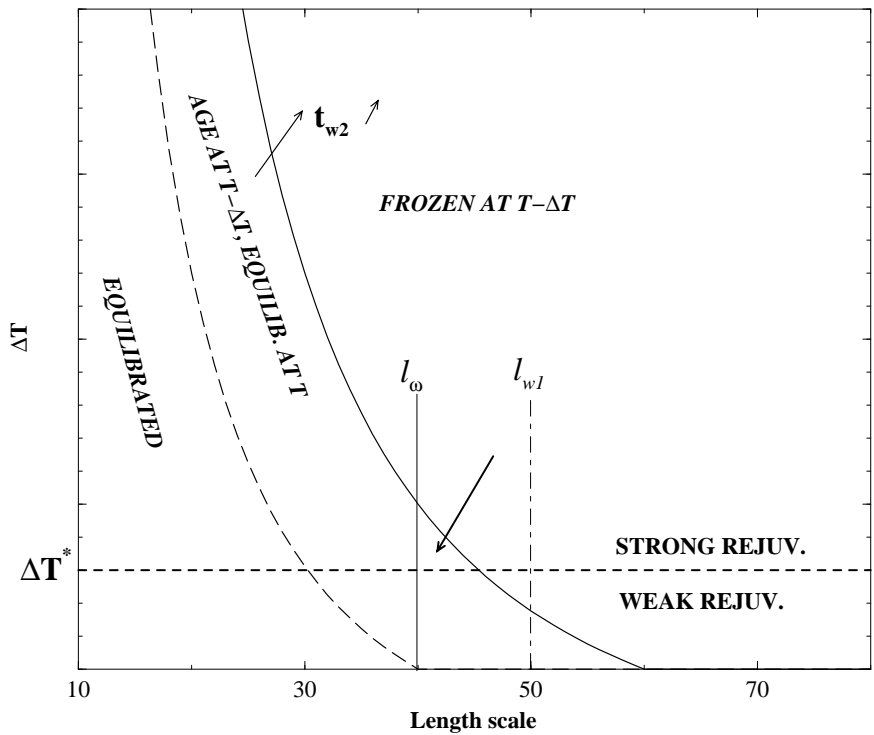,width=10cm,angle=270}
\vskip 0.3cm \caption{\small Schematic view of the plane $\ell,\Delta T$, for a certain time $t_{w1}$ 
spent at $T_1$ (corresponding to length $\ell_1$) and a certain (longer) time spent at 
$T_2=T_1-\Delta T$. For large enough $\Delta T>\Delta T^*$, strong rejuvenation is expected. However, 
very small length scales (below the
dashed curve) are always in equilibrium. Large length scales (above the thick curve, the position of 
which depends on $t_{w2}$) do not evolve at all during the stay at $T_2$. 
Intermediate length scales (between the two curves at to the left of the vertical plain line)
do age at $T_2$, but re-equilibrate at $T_1$ faster than $1/\omega$. These length scales are
responsible for the coexistence of rejuvenation and perfect memory. 
Finally, for large enough $t_{w2}$ or small
enough $\Delta T$, certain length scales (indicated by the arrow) continue to age at $T_2$ and 
re-equilibrate slowly at $T_1$, therefore destroying the perfect memory effect. This is indeed
observed experimentally. A very similar picture would hold in the `chaos' scenario, with the 
constant $\Delta T^*$ line replaced by a crossover line $\Delta T^* \sim \ell^{-1}$.
\label{fig4} }
\end{figure}

\subsection{More sophisticated models}

The simple two-level picture discussed above is obviously not sufficient to explain in details the
experimental data. For example, the aging behaviour of $\chi''$ is well fitted by a power-law:
$\chi''_{AG}(\omega,t_w) \propto (\omega t_w)^{-b}$, with $b \sim 0.2$. This is quite different 
from Eq. (\ref{2levelchi}), although the latter correctly predicts a strong increase of $\chi''$
at low frequencies. On a given length scale $\ell$, there are actually many metastable conformations 
of the domain walls, between which the system jumps. One expects that the energy barriers between
these metastable states are distributed, over a certain energy scale $\Upsilon(T) \ell^\psi$. 
A simple model is to assume that energy barriers are independent random variables -- this is
the trap model \cite{Bouchaud:1992}. There is a limit where this model has a sharply defined behaviour: when 
the distribution
of barriers is exponential, of mean  $T_c(\ell)$, then there is a well defined transition 
temperature which separates an equilibrium phase for $T>T_c(\ell)$ where all configurations 
contribute more or less equally to the equilibrium partition function, from an aging phase for
$T < T_c(\ell)$ where the partition sum is dominated by a few metastable states only \cite{Derrida,
Bouchaud/Mezard,Carpentier/LeDoussal}.
This means that a small temperature change for $T_c(\ell)+\Delta T/2$ to $T_c(\ell)-\Delta T/2$ 
completely changes the way the configuration space is explored. Correspondingly, a strong rejuvenation
effect is expected upon crossing $T_c(\ell)$ (see \cite{Marta} for details). Furthermore, the 
{\it realized} probability to find the system in
a metastable state of lifetime $\tau$ after a waiting time $t_w$ is, for $T < T_c(\ell)$, given
by \cite{Bouchaud:1992}:
\be
P_\ell(\tau,t_w) \sim \frac{t_w^{x_\ell-1}}{\tau^{x_\ell}} {\cal F}\left(\frac{\tau}{t_w}\right)
\qquad x_\ell=\frac{T}{T_c(\ell)} < 1,
\ee
where ${\cal F}(u)$ is a cut-off function, decaying as $1/u$ for large arguments. One can therefore obtain the 
a.c. susceptibility by averaging the simple two-level contribution Eq. (\ref{2levelchi}) over the
realized distribution of trapping times:
\be
\chi''(\omega,t_w)=\int_{\omega^{-1}}^\infty d\tau \  P_\ell(\tau,t_w) \frac{\exp[-t_w/\tau]}{\omega \tau} \propto (\omega t_w)^{-b},
\ee 
with $b=1-x_\ell$ \cite{Balents:1996}. Therefore, the introduction of a 
large number of metastable states allows one 
to obtain both a strong rejuvenation effect at the transition temperature, and a realistic form for
the decay of the a.c. susceptibility. Since there is a continuum hierarchy of length scales, each of 
which corresponding to a different transition temperature $T_c(\ell)$, the problem is actually a multilevel trap
model of the type studied in \cite{Bouchaud/Dean:1995,Nemoto/Sasaki:2000}.
The contributions of the different length scales are intertwined, and one
expects rejuvenation to occur at all temperatures provided the corresponding $\ell$ 
is somewhat larger than the microscopic length. \footnote{The a.c. susceptibility
is however dominated at long times $t_w$ by the length scales such that $x_\ell \sim 1$, therefore explaining 
why $b=1-x_\ell$ 
is found to be small ($1/f$ noise) and nearly temperature independent.}
The coexistence of strong rejuvenation and memory in this model 
has been numerically demonstrated very convincingly in \cite{Nemoto/Sasaki:2000}, and confirm the above
qualitative discussion.

There are many possible variants of the above trap model \cite{Monthus/Bouchaud:1996,Maass:2000}. 
The basic results still hold for more general barrier height distributions, with a weakly time dependent
exponent $x_\ell$ \cite{Monthus/Bouchaud:1996}. One can also consider long-range correlated energy
landscapes, such as the one dimensional Sinai model where barriers typically grow with the position
of the representative point \cite{Bouchaud/Georges:1990}. Many results on aging have been obtained 
for this model \cite{FML}, which has a typical self-similar `valley within valley' structure. Preliminary 
numerical studies also show that rejuvenation and memory effects are also present in the Sinai model
\cite{Sales}. The mechanism at work in these landscape models is very close to the 
hierarchical picture first advocated
by experimentalists to explain rejuvenation and memory \cite{Vincent,Lefloch}: a reduction of temperature 
reveals finer details of the energy landscape within which the system must equilibrate.

\subsection{More on `temperature chaos'}

\subsubsection{Rejuvenation....}

As mentioned in the introduction, the strong rejuvenation effect and the absence of cooling rate
dependence have also been interpreted in terms of `temperature chaos'. Within the droplet model, the
argument suggesting this behaviour is the following: the free energy of an excitation of length $\ell$
is small as a result of the compensation of its energy and entropy, both of them being much larger 
than $\Upsilon \ell^\theta$. Therefore, a small change of temperature $\Delta T$  should ruin 
this subtle compensation on large length scale. This can be interpreted in terms of a 
complete re-shuffling of the dominant 
configuration beyond a certain overlap length $\ell^*$ that diverges when $\Delta T \to 0$. 
The physical mechanism is actually related to the discussion of the previous subsection: since 
the shorter length scales have to reorganize when the temperature changes, the free energy of the
larger length scales will be strongly affected. So the description of large length scales in terms
of simple two (or multi) level systems with a fixed energy landscape is inappropriate: 
the `landscape' itself
is temperature dependent.   

Using the value of the exponents in the droplet model in three
dimensions, one naively estimates $\ell^* \sim \Upsilon(T)/\Delta T$. 
Using the above results on $\Upsilon(T)$,
one finds that for $T=0.7 T_c$, $T_c=15$ K and $\Delta T=1$ K, 
one should have $\ell^* \sim 1$; therefore 
`chaos' effects should be observable both numerically and experimentally. Since no sign of 
chaos was detected numerically for small systems \cite{Billoire}, this suggests that 
numerical prefactors are perhaps large,
and that actually the overlap length is, for practical purposes, larger than the numerically or 
experimentally relevant length scale. This scenario is supported by recent precise numerical studies of
temperature chaos in a simpler problem \cite{YoshinoDP}-- the pinning of a one dimensional domain wall 
(the directed polymer
problem). In this case, the overlap length indeed scales as expected with $\Delta T$, 
although rather large sizes are needed to observe the effect. 
The heuristic scaling arguments for chaos \cite{FH,BM} can furthermore, for this 
simplified 
problem, be made more precise \cite{YoshinoDP}, and suggest that the decorrelation 
beyond the overlap length only decays as a power law rather than an exponential \cite{inprep}. The need to go to very large
length scales appears to be true also for the 2D spin-glass \cite{Huse}. 
Therefore, even if temperature chaos appears to be absent in the 
{\sc sk} model, it is possible that this effect exists in three dimensions, although perhaps on
very large length scales.  

On the other hand, as we discussed above, a hierarchical landscape rejuvenation appears 
sufficient to account for
most of the experimental data. The main distinction is the existence of a characteristic overlap length. In this respect, one reason to believe that `strong' temperature chaos is perhaps not relevant
is that rejuvenation appears to be a small scale, rather than large scale, phenomenon. 
When the temperature is changed sufficiently to induce partial rejuvenation, 
the aging a.c. susceptibility
is made up of two contribution, a {\it short time}, rejuvenation part, and a {\it long time} 
contribution which is the continuation of aging at the first temperature, with a shifted effective
age to account for the modification of time scales with temperature. This suggests that the length
scale $\ell_{1}$ built at $T_1$ actually continues to grow at $T_2$, although by definition, since 
some rejuvenation
is observed, one should be in a regime where $\ell^* < \ell_{1}$. 
The same effect is seen in {\sc zfc}
(zero field cooled) experiments and in numerical simulations \cite{Takayamaexp}: 
after a short transient, the length scale 
that grew at $T_1$ {\it continues to grow} (albeit at a different rate) at $T_2$. 

Another argument against the relevance of an overlap length is provided by the quantitative analysis 
of the effect of a small temperature cycle on the a.c. susceptibility. 
If the length scales $< \ell^*$ are 
unaffected, one expects that the effective initial age $t^*$ of 
the a.c. susceptibility at $T_1 -\Delta T$ is 
such that $t(\ell^*,T_1-\Delta T)=t^*$ (provided that the size of the domains grown at $T_1$ 
is larger than the overlap length, i.e. $\ell_{1} > \ell^*$). We have extracted 
$t^*$ from a set
of experiments where $t_{w1}$ and $T_1$ are fixed, 
and $\Delta T$ is varied. For CrIn, we find that $t^*$ 
behaves as $\Delta T^{-a}$, with $a \sim 1$ for $T/T_c=0.72$ and $a \sim 1.5$ for $T/T_c=0.84$. 
This is incompatible with the 
assumption that $\ell^* \sim \Delta T^{-1}$, which would lead to an extremely fast divergence of 
$t^*$ for small $\Delta T$, according to Eq. (\ref{tcrit}). 
A possibility would be that $\ell^*$ behaves as 
$\Delta T$ to a small negative power, but this violates bounds 
obtained within the framework of the droplet 
model \cite{FH}.

\subsubsection{...and memory ?}

Strong chaos does obviously explain complete rejuvenation for $\Delta T$ sufficiently large. 
In order to be compatible with memory, one should argue, as above, that the length scale $\ell_2$ 
growing at $T_2$ remains somewhat smaller than $\ell_1$; the fast separation of time scales is
then used to account for memory -- see the discussion in 
\cite{Yoshino/Lemaitre:2000}. On the other hand, the magnetic field is known to lead to very strong
`chaos'-like effects. For example, a small magnetic field cycle is sufficient to rejuvenate
completely the system. Furthermore, chaos with magnetic field in the {\sc sk} model was obtained long ago
\cite{ParisiPQ}. An interesting idea is then to perform a simultaneous temperature and magnetic
field cycle, with a $\Delta T$ such that rejuvenation and perfect memory are observed in the
absence of magnetic field changes. In the temperature chaos scenario, adding a magnetic field 
should not change anything since the overlap length is already very small. In the fixed landscape 
scenario, the magnetic field should have an effect since the large scale structures built at $T_1$
will couple to the field (as was discussed in subsection 3.2) and therefore speed up their dynamics at 
$T_2$. Therefore, some loss of memory should be observed, as indeed suggested by the experiments:
see Figure 5. 

\begin{figure}
\hspace*{+1cm}\epsfig{file=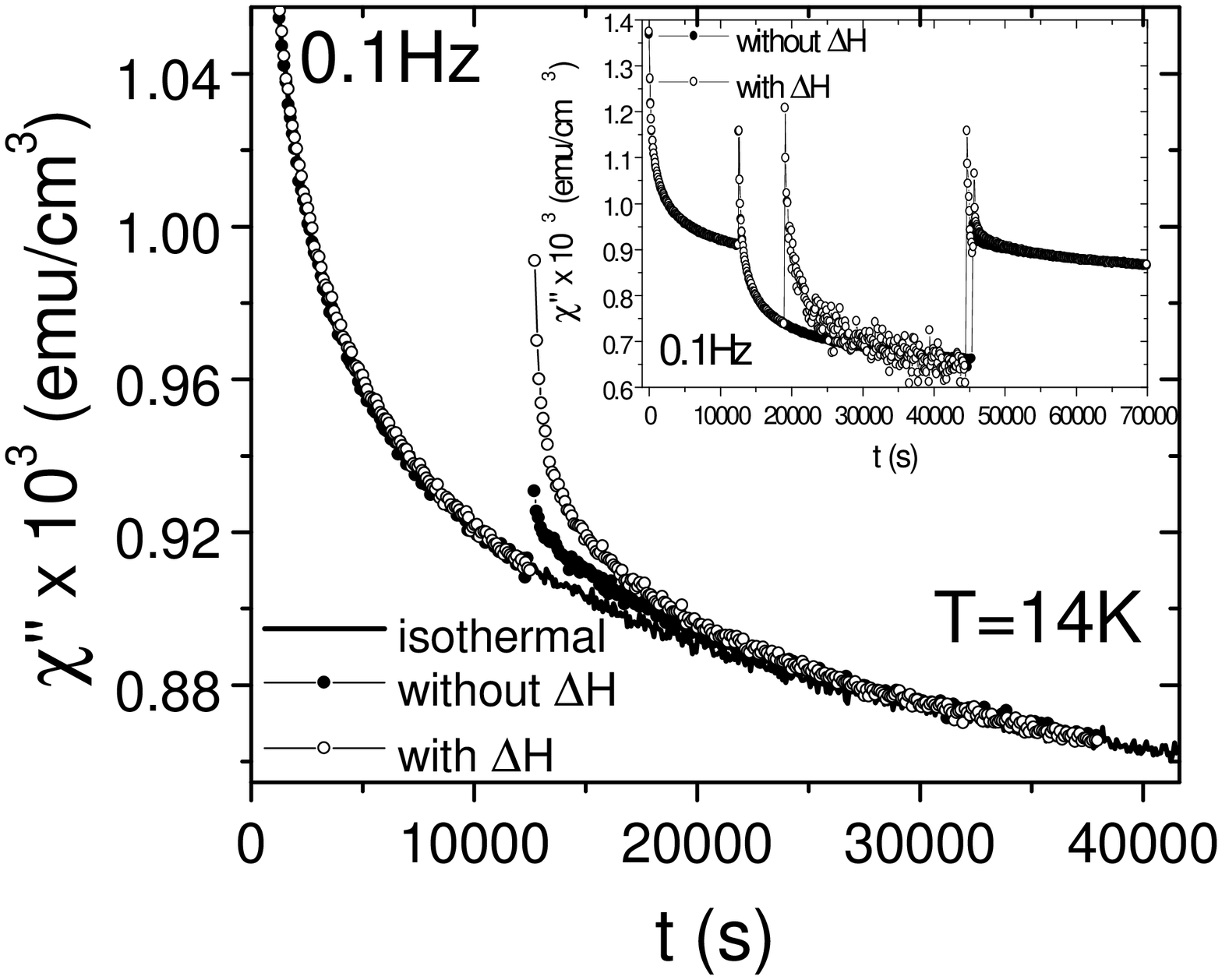,width=10cm}
\vskip 0.3cm \caption{\small Effect of a magnetic field on the memory. Here, the system is cooled from 
$T_1=14$ K to $T_2=12$ K, and heated back to $14$ K. Two experiments have been performed: one where 
the system is unperturbed during the stay at $T_2$, the other where an extra magnetic field of 
$\Delta H=60$ 
Gauss is imposed. This magnetic field is known to lead to a strong rejuvenation effect. 
The choice of parameters is such that memory is not perfect, even for $\Delta H=0$. Here, we see
that the effect of $\Delta H$ is noticable, which shows that the purely thermal overlap length
cannot be small.    
\label{fig5} }
\end{figure}

Following this line of thought, we have also performed an experiment to test a
very spectacular prediction of \cite{Yoshino/Lemaitre:2000}: a double temperature 
cycling where $T_1 \to T_2=T_1 -\Delta T$
is followed, after a certain time $t_{w2}$ at $T_2$ by a second small quench 
$T_2 \to T_3=T_2 -\Delta T'$. If the 
system is re-heated rapidly to $T_1$ without stopping at $T_2$, 
the dramatic prediction of \cite{Yoshino/Lemaitre:2000}
is that the memory effect should be destroyed. 
Only if the system is allowed to `take its breath' at $T_2$ will memory be
preserved. The problem is that `re-heating rapidly' 
cannot be achieved experimentally, since the fastest achievable temperature
ramps allows the system to stay at each temperature during a time which is huge in 
microscopic units. This might be
enough (again due to the fast separation of time scales) to allow for memory conservation ! 
Our idea was then to use
the magnetic field to prevent the system from tracing back its previous history, which is 
the crucial ingredient, in the chaos scenario, to preserve the memory (see the
detailed discussion in  \cite{Yoshino/Lemaitre:2000}). 
We have therefore applied a magnetic field {\it while the system is 
re-heated}: the target state at $T_2$ is therefore completely scrambled. 
In spite of this, the loss of memory when returning at 
$T_1$ is the same as for a single quench procedure. 
Therefore, the spectacular effect predicted within the `strong chaos' 
scenario fleshed out in \cite{Yoshino/Lemaitre:2000} is not observed. 

 \begin{figure}
\hspace*{+1cm}\epsfig{file=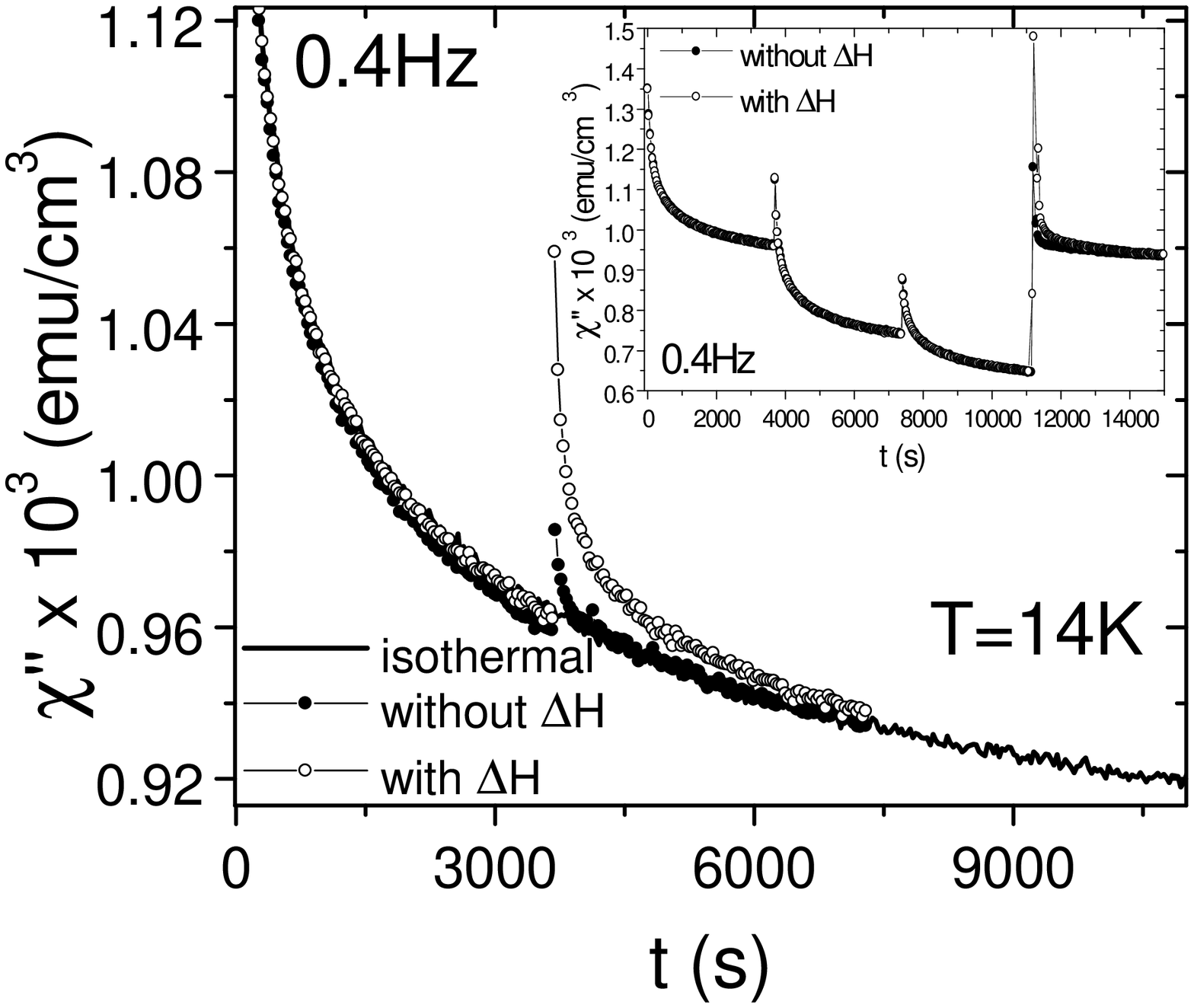,width=10cm}
\vskip 0.3cm \caption{\small Now, the system is cooled from 
$T_1=14$ K to $T_2=12$ K, and then to $T_3=10$ K in zero field. When the system is 
heated back, an extra magnetic field is imposed between $11$ K and $13$ K. 
In the `strong' chaos scenario, this should prevent the system from `remembering'
its previous $12$ K history, and should completely scramble the memory effect at $14$ K.
The comparison between this data and that of Figure 5 shows that the loss of memory is only partial,
and nearly identical in the two cases. Therefore, the spectacular loss of memory in a double
quench experiment suggested in \cite{Yoshino/Lemaitre:2000}, is not observed.      
\label{fig6} }
\end{figure}

Finally, note that in the chaos scenario, it should be possible to 
choose $\Delta T$, $t_{w1}$ and $t_{w2}$ 
in a {\it positive} temperature cycling such that $\ell_2 < \ell_1$. In this case, memory should be
preserved in a positive cycling experiment, {\it even if some rejuvenation takes place} at $T_2$. 
To our knowledge, this has never been observed, but this
might again be due to the fact that the time scales separate very quickly. 
Taking the same value of parameters
as in Fig. 2, we find that for $T_2=0.77 T_c$, $t_{w2}=10$ sec., $\ell_2 \sim 20$. One would need to
wait at least $t_{w1} \sim 10^5$ seconds (1.5 day) at $T_1=0.7 T_c$ to reach the same length and
allow memory to be preserved. 
Note that the fundamental assymmetry between small positive and small negative
temperature jumps is probably  the most clear cut difference between the `chaos' scenario and the hiererchical landscape scenario. It would therefore be crucial to find an experimental situation 
where rejuvenation and memory in a positive $\Delta T$ cycle should in principle be observed.
It would be interesting to study this issue in the Ising-like 
sample studied in \cite{Ising}, where the separation of time scales is milder.

\section{Conclusion -- Open problems}

In this paper, we have summarized the different puzzles raised by aging experiments 
of spin-glasses and their different interpretations.  We try to reconcile the  `real space', droplet
like pictures and the hierarchical pictures that have been proposed in the past. The
basic ingredient is a strong separation 
of the time scales that govern the dynamics of the system on different length scales. Changing
the temperature changes the length scale at which the system is observed, thereby allowing 
rejuvenation (that concerns short length scales) and memory (stored in long length scales) to
coexist. We have shown that previous experiments can be reanalyzed in terms of {\it vanishing
energy barriers} at the spin-glass transition, which in turn leads to the `super-activated' 
behaviour observed in several experiments. We have argued that the power-law dependence
of the coherence length with time might actually reflect a slow crossover form critical to
activated dynamics. Finally, we have tried to distinguish between a hierarchical 
landscape rejuvenation, which is already 
present in simple multi-level systems like the Random Energy Model 
\cite{Marta}, 
from a more sophisticated 
`strong' chaos that arises because large length scales free energies are renormalized by
small length scale fluctuations \cite{BM,FH}. We have argued that most experiments can be 
accounted for 
without invoking the existence of an overlap length. Some specific predictions of the 
strong chaos scenario have been 
tested and our not borne out by our new experimental results. 
Nevertheless, we believe that this effect should exist on 
sufficiently large length scales, but these are perhaps out of reach 
both from numerical and experimental 
possibilities. It should finally be noted that 
rejuvenation and memory effects have also been observed 
in other, very different systems, such as {\sc pmma} \cite{pmma}, where the relevance 
of temperature chaos is not clear, whereas a scenario based on multiscale dynamics is plausible.

From a theoretical point of view, the dynamics of mean field models corresponding to full
replica symmetry breaking has been shown to exhibit rejuvenation and memory effects \cite{CuKuRejMem},
and are actually very closely related to models of diffusion in self-similar landscapes 
such as the Sinai model \cite{CuLeDoussal}, although the precise role of activation in 
these models is still rather obscure. However, these models are in principle incompatible 
with simple $t/t_w$ aging, as experimentally \cite{RevSitges} and numerically \cite{Picco} observed. 
It would be gratifying to understand whether or not these mean field models can 
correctly be interpreted in finite dimensions in terms of the simple trap models \cite{Balents:1996}.
This question of course has a far more general scope and relevant for other glassy systems as well.  

\vskip 1cm

{\it Acknowledgements} We are indebted to H. Yoshino for sharing with us many interesting ideas. 
We want to thank L. Berthier, E. Bertin, P. Doussineau, I. Giardina, J. Kurchan, 
J. Lamarcq, A. Levelut, O. Martin, M. M\'ezard and H. Takayama for interesting discussions. 
J.-P. B. wants 
to thank Harvard University for hospitality where part of this work was performed.  
Very stimulating discussions with D.S. Fisher, J.P. Sethna and R. da Silveira 
have to be acknowledged. This work
owes a lot to the inspiring ideas of Daniel Fisher.

\end{document}